\documentclass[preprints,article,accept,moreauthors,pdftex,10pt,a4paper]{mdpi}

\graphicspath{{figures/}}

\firstpage{1}
\pubvolume{xx}
\issuenum{1}
\articlenumber{1}
\pubyear{2018}
\copyrightyear{2018}
\history{
    Preprint uploaded XXXXXX
    Received: date;
    Accepted: date;
    Published: date
}

\pdfoutput=1

\Title{Actuator line modeling of vertical-axis turbines}

\Author{
    Peter Bachant $^{1}$\orcidA{},
    Anders Goude $^{2}$
    and Martin Wosnik $^{1,}$*
}

\AuthorNames{Peter Bachant, Anders Goude and Martin Wosnik}

\address{%
$^{1}$ \quad University of New Hampshire, USA\\
$^{2}$ \quad Uppsala University, Sweden}

\corres{Correspondence: martin.wosnik@unh.edu}

\abstract{
    To bridge the gap between high and low fidelity numerical modeling tools for
    vertical-axis (or cross-flow) turbines (VATs or CFTs), an actuator line
    model (ALM) was developed and validated for both a high and a medium
    solidity vertical-axis turbine at rotor diameter Reynolds numbers $Re_D \sim
    10^6$. The ALM is a combination of classical blade element theory and
    Navier--Stokes based flow models, and in this study both $k$--$\epsilon$
    Reynolds-averaged Navier--Stokes (RANS) and Smagorinsky large eddy
    simulation (LES) turbulence models were tested using the open-source
    OpenFOAM computational fluid dynamics framework. The RANS models were able
    to be run on coarse grids while still providing good convergence behavior in
    terms of the mean power coefficient, and also approximately four orders of
    magnitude reduction in computational expense compared with 3-D
    blade-resolved RANS simulations. Submodels for dynamic stall, end effects,
    added mass, and flow curvature were implemented, resulting in reasonable
    performance predictions for the high solidity rotor, more discrepancies for
    the medium solidity rotor, and overprediction for both cases at high tip
    speed ratio. The wake results showed that the ALM was able to capture some
    of the important flow features that contribute to VAT's relatively fast wake
    recovery---a large improvement over the conventional actuator disk model.
    The mean flow field was better realized with the LES, which still
    represented a computational savings of two orders of magnitude compared with
    3-D blade-resolved RANS, though vortex breakdown and subsequent turbulence
    generation appeared to be underpredicted, which necessitates further
    investigation of optimal subgrid scale modeling.
}


\begin{document}



\section{Introduction}

Vertical-axis (cross-flow) turbines (VATs or CFTs) were the subject of
significant research and development in the 1970s through the 1990s by groups
like Sandia National Labs in the US~\cite{Sutherland2012} and the National
Research Council of Canada~\cite{Para2002}. Despite minor commercial success for
large scale onshore wind applications, vertical-axis turbines were virtually
abandoned in favor of horizontal-axis (or axial-flow) turbines, as they
generally are more efficient and do not encounter the high levels of fatigue
loading that VATs do. Today, however, there is renewed interest in VATs for
marine hydrokinetic (MHK) applications~\cite{ORPC2012}, offshore floating wind
farms~\cite{Paulsen2011, Sandia2012, Dodd2014}, and smaller scale, tightly
spaced wind farms~\cite{Dabiri2011, Kinzel2012}, thanks to their relatively
faster wake recovery.


The mean near-wake structure of vertical-axis turbine has been shown to be
largely dominated by the effects of tip vortex shedding, which induces levels of
recovery due to vertical advection significantly larger than those from
turbulent fluctuations~\cite{Bachant2015-JoT}---an effect not seen in axial-flow
or horizontal-axis turbine (AFT or HAT) wakes. As interest shifts to designing
and analyzing arrays of VATs, it is necessary to determine the effectiveness of
various numerical modeling techniques to replicate VAT near-wake dynamics, such
that wake recovery is accurately predicted, leading to accurate assessment of
optimal array spacing.

Reynolds-averaged Navier--Stokes (RANS) turbulence models computed on 3-D
body-fitted (blade-resolved) grids can do a good job predicting the mean
performance and near-wake structure of a VAT, though their effectiveness depends
on the turbulence model
applied~\cite{Bachant2016-BR-CFD,Lam2016,Alaimo2015,Boudreau2017,Marsh2015}.
However, 3-D blade-resolved RANS presents a huge computational expense---on the
order of 1,000 CPU hours per simulated second with contemporary hardware---since
it must resolve fine details of the blade boundary layers, which will preclude
its use for array analysis until the availability of computing power increases
sufficiently. It is therefore necessary to explore simpler models that can
predict the turbine loading and flow field with acceptable fidelity, but that
are economical enough to not require high performance computing, at least for
individual devices.


For analyzing turbine arrays, it is desirable to retain a Navier--Stokes
description of the flow field---in contrast to, e.g., momentum or potential flow
vortex models---to capture the effects of nonlinear advection and turbulent
transport. However, rather than resolving the fine details in the blade boundary
layers, an actuator-type model for parameterizing the turbine loading may be
employed, which dramatically drives down computational expense. As shown in
\cite{Bachant2015-JoT}, the conventional uniform actuator disk is not a good
candidate for a cross-flow turbine wake generator, never mind the fact that it
does not typically compute performance predictions.

Actuator line modeling (ALM), originally developed by Sorensen and Shen
\cite{Sorensen2002}, is an unsteady method that tracks blade element locations,
and has become popular for modeling axial-flow or horizontal-axis turbines, and
has been shown in blind tests to be competitive with blade-resolved
CFD~\cite{Krogstad2013, Pierella2014}. The ALM combined with large eddy
simulation (LES) has become the state-of-the-art for modeling entire wind
farms~\cite{Archer2013, Churchfield2012, Sorensen2015, Fleming2013,
Fleming2014}. Like other blade element techniques, the effectiveness of the ALM
for AFTs is in part due to the quasi-steady nature of the flow in the blade
reference frame, and the relatively rare occurrence of stall. Note that a
similar method can be used with the Navier--Stokes equations in
vorticity--velocity form~\cite{Scheurich2011b}, which may be more efficient when
vorticity is confined to a relatively small fraction of the domain, e.g., for a
standalone turbine with a uniform laminar inflow.


The ALM has been previously used to model a very low Reynolds number (based on
rotor diameter, $Re_D \sim 10^4$) 2-D cross-flow turbine experiment in a flume
using large eddy simulation (LES)~\cite{Shamsoddin2014}. Performance predictions
for this case were not reported, but the ALM was shown to be more effective at
postdicting the wake characteristics measured in the experiments by Brochier et
al.~\cite{Brochier1986}.

\begin{figure}
    \centering

    \includegraphics[width=0.6\textwidth]{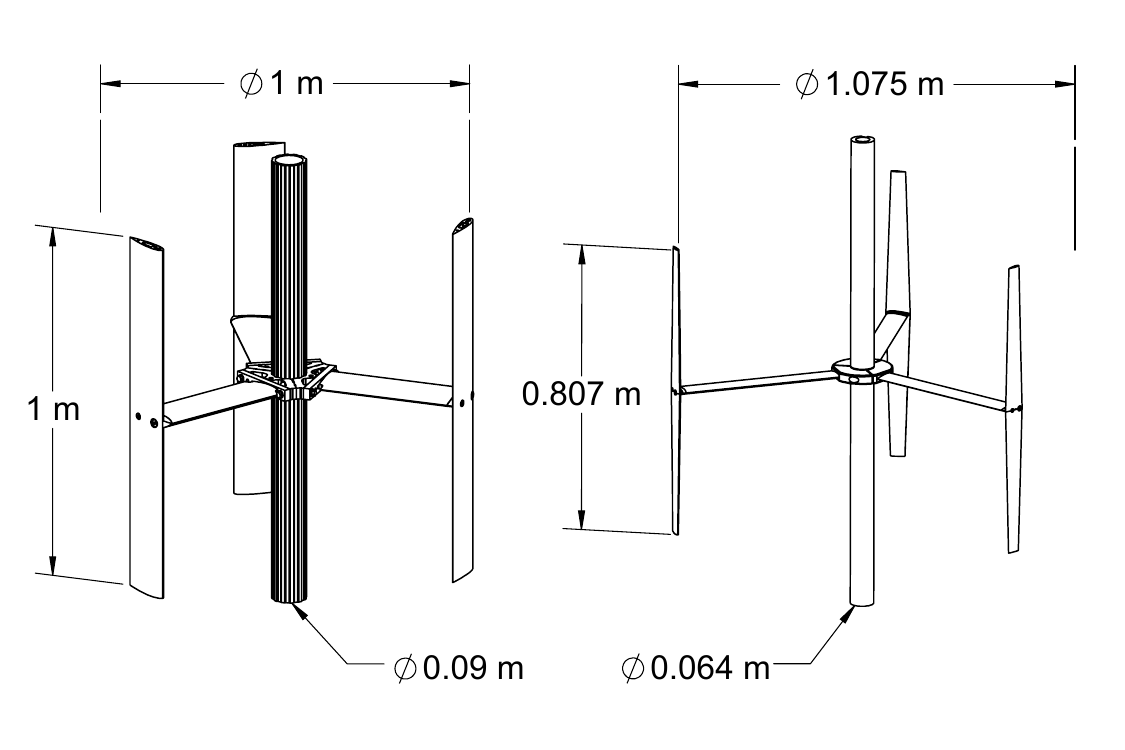}

    \caption{Drawings of the turbines modeled in this study: the UNH-RVAT (left)
    and DOE/SNL RM2 (right).}

    \label{fig:turbines}
\end{figure}

In this study we have developed an ALM for cross-flow turbines and embedded the
model inside LES and unsteady RANS simulations, the latter of which has not yet
been reported in the literature. This model was implemented within the
open-source OpenFOAM CFD framework and validated against experimental data for
the high solidity ($c/R=0.28$) UNH Reference Vertical-Axis Turbine (UNH-RVAT)
and the medium solidity ($c/R=0.07$--$0.12$) US Department of Energy/Sandia
National Labs Reference Model 2 (RM2) cross-flow turbine (at 1:6 scale), both of
which are shown in Figure~\ref{fig:turbines}. Validation datasets for both
turbines were taken from \cite{Bachant2016-RVAT-Re-dep} and
\cite{Bachant2016-RM2-data}, respectively. The turbines were modeled at $Re_D
\sim 10^6$, at which experimentally measured performance and near-wake
characteristics were nearly Reynolds number
independent~\cite{Bachant2016-Energies,Bachant2016-RM2-paper}, and two orders of
magnitude higher than previous ALM investigations for VATs.


\section{Theory}

\begin{figure}
    \centering

    \includegraphics[clip, trim=1in 1.5in 1in 0.5in,
    width=0.4\textwidth]{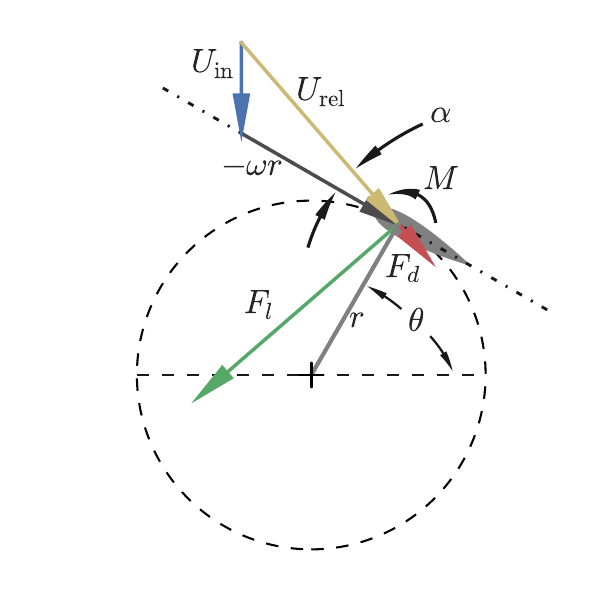}

    \caption{Vector diagram of velocity and forcing on a cross-flow turbine
    blade element. Note that the free stream velocity $U_\infty$ is oriented
    from top to bottom (identical to $U_{\mathrm{in}}$ for purely geometric
    calculations), the blade chord (dash-dotted line) is coincident with the
    tangential velocity (i.e., zero preset pitch, which would offset the
    geometric angle of attack $\alpha$), and the drag vector is magnified by a
    factor of two (approximately, relative to the lift vector) to enhance
    visibility.}

    \label{fig:vectors}
\end{figure}

The actuator line model is based on the classical blade element theory combined
with a Navier--Stokes description of the flow field. The ALM treats turbine
blades as lines of blade or actuator line elements, defined by their
quarter-chord location, and for which 2-D profile lift and drag coefficients are
known. For each blade element, relative flow velocity $\vec{U}_\mathrm{rel}$ and
angle of attack $\alpha$ are computed by adding the vectors of relative blade
motion $-\omega r$, where $\omega$ is the rotor angular velocity and $r$ is the
blade element radius, and the local inflow velocity $\vec{U}_\mathrm{in}$, a
diagram of which is shown in Figure~\ref{fig:vectors}.


Inflow velocity was sampled for each actuator line element at its quarter-chord
location using OpenFOAM's \texttt{interpolationCellPoint} class, which provides
a linear weighted interpolation using cell values. This algorithm helps keep the
sampled velocity ``smooth'' compared with using the cell values themselves,
especially when elements are moving in space as they are in a turbine, since
meshes will likely have a cell size on the same order as the chord length, and
will move on the order of one cell length per time step.

The kinematics of a CFT are parameterized by the tip speed ratio $\lambda =
\omega R/U_\infty$, where $R$ is the maximum rotor radius, and $U_\infty$ is the
free stream velocity. In contrast to an AFT, which under ideal conditions can be
considered a steady machine, CFT blades encounter large amplitude and rapid
oscillations in angle of attack, often exceeding the rotor blades' static stall
angles in typical operation, which makes their blade loading and performance
more difficult to predict. The unsteadiness can be characterized by a reduced
frequency~\cite{Leishman2006}
\begin{equation}
    k = \frac{\omega c}{2 U_\infty},
\end{equation}
which assumes the free stream velocity is constant.
Unsteady effects begin to
become significant for $k > 0.05$, and can become dominant for $k \ge 0.2$.
Note $\omega$ in the reduced frequency equation refers to the frequency of
angle of attack oscillation, which is normally sinusoidal in pitching foil
experiments, but is analogous to cross-flow turbine blade kinematics, i.e.,
the angle of attack cycle encounted by a CFT blade has a period equal to one
turbine revolution.


For a VAT or CFT reduced frequency can be reformulated in terms of the tip speed
ratio as
\begin{equation}
    k = \frac{\lambda c}{2R},
\end{equation}
which is then
also a function of solidity or chord-to-radius ratio $c/R$. As an example, a
large scale, relatively low solidity Darrieus wind turbine such as the Sandia 34
m diameter Test Bed, with an equatorial blade chord of 0.91 m \cite{Murray2011},
a reduced frequency $k=0.16$ is encountered based solely on angle of attack
oscillations at $\lambda=6$. For the two turbines studied here, reduced
frequencies at the tip speed ratio of peak power coefficient, $\lambda_0$,
are 0.27 and 0.17 (based on root chord length) for
the UNH-RVAT and RM2, respectively. Therefore, both cases will be inherently
unsteady, with these effects dominating the UNH-RVAT's behavior. Hence, an
accurate unsteady aerodynamics model is important to predicting blade loading.

Assuming unsteady effects can be appropriately modeled, the blade element lift
force, drag force, and pitching moment are calculated as
\begin{equation}
    F_l = \frac{1}{2} \rho A_\mathrm{elem} C_l |\vec{U}_\mathrm{rel}|^2,
\end{equation}
\begin{equation}
    F_d = \frac{1}{2} \rho A_\mathrm{elem} C_d |\vec{U}_\mathrm{rel}|^2,
\end{equation}
\begin{equation}
    M = \frac{1}{2} \rho A_\mathrm{elem} c C_m |\vec{U}_\mathrm{rel}|^2,
\end{equation}
respectively, where $\rho$ is the fluid density, $A_\mathrm{elem}$ is the blade
element planform area (span $\times$ chord), $\vec{U}_\mathrm{rel}$ is the local
relative velocity projected onto the plane of the element profile cross-section
(i.e., the spanwise component is neglected), and $C_l$, $C_d$, and $C_m$ are the
sectional lift, drag, and pitching moment coefficients, respectively, which are
linearly interpolated from a table per the local angle of attack. The forces are
then projected onto the rotor coordinate system to calculate torque, overall
drag, etc. Forces from the turbine shaft and blade support struts are computed
in a similar way. After the force on the actuator lines from the flow is
computed, it is then added to the Navier--Stokes equations as a body force or
momentum source (per unit density, assuming incompressible flow):
\begin{equation}
    \frac{\mathrm{D} \vec{u}}{\mathrm{D} t} = - \frac{1}{\rho} \nabla p + \nu
    \nabla^2 \vec{u} + F_\mathrm{turbine}.
\end{equation}


\section{Static profile coefficient data}

Static input foil coefficient data were taken from Sheldahl and
Klimas~\cite{Sheldahl1981}---a popular database developed for CFTs, which
contains values over a wide range of Reynolds numbers. The Sheldahl and Klimas
dataset has some limitations, namely that data for some foil and/or Reynolds
numbers were ``synthesized'' numerically from other measurements. Despite its
flaws, this dataset is the likely most comprehensive available with respect to
variety of profiles and ranges of Reynolds numbers. Surprisingly, considering
the maturity and popularity of NACA foils, data remains scarce, especially for
$Re_c \sim 10^5$.


NACA 0021 coefficients were used for both turbines, despite the fact that the
UNH-RVAT is constructed from NACA 0020 foils, as a NACA 0020 dataset was not
available---it is assumed the small difference in foil thickness is negligible.
Since pitching moment data were only available at limited Reynolds numbers, two
datasets were used: The lowest for $Re_c \leq 3.6 \times 10^5$ and highest $Re_c
\geq 6.8 \times 10^5$. For each actuator line element, blade chord Reynolds
number is computed based on the sampled inflow velocity, and the static
coefficients are then interpolated linearly within the database.

Each rotor's shaft was assumed to have a drag coefficient $C_d = 1.1$, and the
blade support strut end element drag coefficients were set to 0.05, to
approximate the effects of separation in the corners of the blade--strut
connections.

\section{Force projection}

After the force on the actuator line element from the flow is calculated,
it is then projected
back onto the flow field as a source term in the momentum equation. To avoid
instability due to steep gradients, the source term is tapered from its maximum
value away from the element location by means of a spherical Gaussian function.
The width of this function $\eta$ is controlled by a single parameter
$\epsilon$, which is then multiplied by the actuator line element force and
imparted on a cell with distance $| \vec{r} |$ from the actuator line element
quarter chord location:
\begin{equation}
    \eta = \frac{1}{\epsilon^3 \pi^{3/2}} \exp
    \left[ - \left( \frac{| \vec{r} |}{\epsilon} \right)^2 \right].
    \label{eq:projection}
\end{equation}

Troldborg~\cite{Troldborg2008} proposed that the Gaussian width should be set to
twice the local cell length $\Delta x$ in order to maintain numerical stability.
Schito and Zasso \cite{Schito2014} found that a projection $\epsilon$ equal to
the local mesh length was optimal. Jha et al.~\cite{Jha2014} investigated the
ideal projection width for HAWT blades, recommending an equivalent elliptic
planform be constructed and used to calculate a spanwise $\epsilon$
distribution.

Martinez-Tossas and Meneveau \cite{Martinez-Tossas2015b} used a 2-D potential
flow analysis to determine that the optimal projection width for a lifting
surface is 14--25\% of the chord length. The width due to the wake caused by the
foil drag force was recommended to be on the order of the momentum thickness
$\theta$, which for a bluff body or foil at large angle of attack is related to
the drag coefficient ($O(1)$) by \cite{TennekesAndLumley}
\begin{equation}
    C_d = 2 \theta / l,
    \label{eq:mom-thickness}
\end{equation}
where $l$ is a reference length, e.g., diameter for a cylinder or chord length
for a foil.

Using these guidelines, three Gaussian width values were determined: one
relative to the chord length, one to the mesh size, and one to the momentum
thickness due to drag force. Each three were computed for all elements at each
time step, and the largest was chosen for the force projection algorithm. Using
this adaptive strategy, fine meshes could benefit from the increased accuracy of
more concentrated momentum sources, and coarse meshes would be protected from
numerical instability.

The Gaussian width due to mesh size $\epsilon_{\mathrm{mesh}}$ was determined
locally on an element-wise basis by estimating the size of the cell containing
the element as
\begin{equation}
    \Delta x \approx \sqrt[3]{V_\mathrm{cell}},
\end{equation}
where $V_\mathrm{cell}$ is the cell volume. To account for the possibility of
non-unity aspect ratio cells, an additional factor $C_\mathrm{mesh}$ was
introduced, giving
\begin{equation}
    \epsilon_{\mathrm{mesh}} = 2C_\mathrm{mesh} \Delta x.
\end{equation}
$C_{\mathrm{mesh}}$ was set to 2.0 for the simulations presented
here---determined by trial-and-error to provide stability on the finest grids.
However, in the ALM code $C_{\mathrm{mesh}}$ is selectable at run time for each
profile used.

\section{Unsteady effects}

In the context of a turbine---especially a cross-flow turbine---the actuator
lines will encounter unsteady conditions, both in their angle of attack and
relative velocity. These conditions necessitate the use of unsteady aerodynamic
models to augment the static foil characteristics, both to capture the time
resolved response of the attached flow loading and effects of flow acceleration,
also known as added mass. Furthermore, the angles of attack encountered by a CFT
blade will often be high enough to encounter dynamic stall (DS). It is therefore
necessary to model both unsteady attached and detached flow to obtain accurate
loading predictions.

\subsection{Dynamic stall}

In this study we employed a dynamic stall model developed for low mach numbers
by Sheng et al. \cite{Sheng2008}---derived from the Leishman--Beddoes (LB)
semi-empirical model \cite{Leishman1989}. This model, along with two other
Leishman--Beddoes model variants, was tested for its effectiveness in cross-flow
turbine conditions by Dyachuk et al.~\cite{Dyachuk2014}, who concluded that the
Sheng et al. variant results matched most closely with experiments. In a similar
study \cite{Dyachuk2015}, the Sheng et al. model also performed better than the
Gormont model \cite{Gormont1973}, which inspired its adoption here for the ALM.
Note that even in the absence of stall, the LB DS models still modulate foil
force and moment coefficients to account for unsteadiness, which is important
for the reduced frequencies of the rotors simulated here.

Before the dynamic stall subroutine is executed, the static profile data for
each element is interpolated linearly based on local chord Reynolds number,
which is calculated based on the detected inflow velocity for each actuator
line element and the simulation's kinematic viscosity. The
profile data characteristics---static stall angle, zero-lift drag coefficient,
and separation point curve fit parameters---are then recomputed each time step
such that the effects of Reynolds number on the static data are included.

Inside the ALM, angle of attack is sampled from the flow field rather than
calculated based on the geometric angle of attack. Therefore, the implementation
of the LB DS model was such that the equivalent angle of attack
$\alpha_\mathrm{equiv}$ was taken as the sampled rather than the lagged
geometric value. A similar implementation was used by Dyachuk et al.
\cite{Dyachuk2015a} inside a vortex model.

\subsection{Added mass}

A correction for added mass effects, or the effects due to accelerating the
fluid, was taken from Strickland et al.~\cite{Strickland1981}, which was derived
by considering a pitching flat plate in potential flow. In the blade element
coordinate system, the normal and chordwise (pointing from trailing to leading
edge, which is opposite the $x$-direction used by Strickland et al.)
coefficients due to added mass are
\begin{equation}
    C_{n_\mathrm{AM}} = -\frac{\pi c \dot{U_n}}{8 | U_\mathrm{rel} |^2},
\end{equation}
and
\begin{equation}
    C_{c_\mathrm{AM}} = \frac{\pi c \dot{\alpha} U_n }{8 | U_\mathrm{rel} |^2},
\end{equation}
respectively, where $U_n$ is the normal component of the relative velocity, and
dotted variables indicate time derivatives, which were calculated using a simple
first order backward finite difference. Similarly, the quarter-chord moment
coefficient due to added mass was calculated as
\begin{equation}
    C_{m_\mathrm{AM}} = -\frac{C_{n_\mathrm{AM}}}{4}
        - \frac{U_n U_c}{8 | U_\mathrm{rel} |^2},
\end{equation}
where $U_c$ is the chordwise component of relative velocity. Note that the
direction of positive moment is ``nose-up,'' which is opposite that used by
Strickland et al.

The normal and chordwise added mass coefficients translate to lift and drag
coefficients by
\begin{equation}
    C_{l_\mathrm{AM}} = C_{n_\mathrm{AM}} \cos \alpha + C_{c_\mathrm{AM}} \sin
    \alpha,
\end{equation}
and
\begin{equation}
    C_{d_\mathrm{AM}} = C_{n_\mathrm{AM}} \sin \alpha - C_{c_\mathrm{AM}} \cos
    \alpha,
\end{equation}
respectively. The added mass coefficients were then added to those calculated by
the dynamic stall model.

\section{Flow curvature corrections}

The rotating blades of a cross-flow turbine will have non-constant chordwise
angle of attack distributions due to their circular paths---producing so-called
flow curvature effects~\cite{Migliore1980}. This makes it difficult to define a
single angle of attack for use in the static coefficient lookup tables.
Furthermore, this effect is more pronounced for high solidity ($c/R$) turbines.

The flow curvature correction used here was derived in Goude \cite{Goude2012} by
considering a flat plat moving along a circular path in potential flow, for
which the effective angle of attack including flow curvature effects is given by
\begin{equation}
    \alpha = \delta + \arctan \frac{V_\mathrm{abs} \cos(\theta_b -
        \beta)}{V_\mathrm{abs} \sin(\theta_b - \beta) + \Omega R} - \frac{\Omega
        x_{0r}c}{V_\mathrm{ref}} - \frac{\Omega c}{4 V_\mathrm{ref}},
    \label{eq:Goude-curvature}
\end{equation}
where $\delta$ is the blade pitch angle, $V_\mathrm{abs}$ is the magnitude of
the local inflow velocity at the blade, $\theta_b$ is the blade azimuthal
position, $\beta$ is the direction of the inflow velocity, $\Omega$ is the
turbine's angular velocity, $R$ is the blade element radius, $x_{0r}$ is a
normalized blade attachment point along the chord (or fractional chord distance
of the mounting point from the quarter-chord), $c$ is the blade chord length,
and $V_\mathrm{ref}$ is the reference flow velocity for calculating angle of
attack.

In the actuator line model, each element's angle of attack is calculated using
vector operations, which means the first two terms in
Equation~\ref{eq:Goude-curvature} are taken care of automatically since each
element's inflow velocity, chord direction, and element velocity vectors are
tracked. Therefore, the last two terms in Equation~\ref{eq:Goude-curvature} were
simply added to the scalar angle of attack value. Note that for a cross-flow
turbine, this correction effectively offsets the angle of attack, which
therefore increases its magnitude on the upstream half of the blade path, and
decreases its magnitude on the downstream half, where the angle of attack is
negative.

\section{End effects}

Helmholtz's second vortex theorem states that vortex lines may not end in a
fluid, but must either form closed loops or extend to boundaries. Consequently
the lift distribution due to the bound vortex from foils of finite span must
drop to zero at the tips. One popular end effects correction was developed by
Glauert~\cite{Glauert1935} for the blade element analysis of axial-flow rotors.
However this correction depends on rotor parameters---tip speed ratio, number of
blades, element radius, tip flow angle---that do not necessarily translate
directly to the geometry and flow environment of a cross-flow rotor. Therefore,
a more general end effects model was sought.

From Prandtl's lifting line theory, the geometric angle of attack $\alpha$ of a
foil with an arbitrary circulation distribution can be expressed as a function
of nondimensional span $\theta$ as \cite{Anderson2001}
\begin{equation}
    \alpha (\theta) = \frac{2S}{\pi c (\theta)}
    \sum_1^N A_n \sin \theta
    + \sum_1^N n A_n \frac{\sin n \theta}{\sin \theta}
    + \alpha_{L = 0}(\theta),
    \label{eq:lifting-line}
\end{equation}
where $S$ is the span length, $c(\theta)$ is the chord length, and $N$ is the
number of locations or elements sampled along the foil. This relationship can be
rearranged into a matrix equation to solve for the unknown Fourier coefficients
$A_n$, after which the circulation distribution can be calculated as
\begin{equation}
    \Gamma (\theta) = 2SU_\infty \sum_1^N A_n \sin n \theta,
\end{equation}
which, via the Kutta--Joukowski theorem, provides the lift coefficient
distribution
\begin{equation}
    C_l(\theta) = \frac{-\Gamma (\theta)}{\frac{1}{2} c U_\infty}.
\end{equation}

We can therefore compute a correction function $F$ to be applied to the ALM lift
coefficient, based on the normalized spanwise lift coefficient distribution
\begin{equation}
    F = C_l(\theta)/C_l(\theta)_{\max},
\end{equation}
which will be in the range $[0, 1]$, similar to the Glauert corrections, but
does not contain rotor parameters.

\section{Software implementation}

NREL has developed and released an actuator line modeling library,
SOWFA~\cite{Churchfield2014b}, for simulating horizontal-axis wind turbine
arrays using the OpenFOAM finite volume CFD library. OpenFOAM is free,
open-source, widely used throughout industry and academia, and has grown a very
active support community around itself. Though SOWFA is also open-source, its
architecture is focused on simulating horizontal-axis wind turbine arrays
in the atmospheric boundary layer and would have required significant effort to
adapt for cross-flow turbines. Thus, a new and more general ALM library
was developed that could model both cross- and axial-flow
turbines, as well as standalone actuator lines. The actuator line model
developed here, turbinesFoam \cite{Bachant2018-turbinesFoam-v0.0.8}, was
also written as an extension library for OpenFOAM (v5.0), using its
\texttt{fvOptions} framework for adding source terms to equations at runtime.
This implementation allows the CFT-ALM to be added to many of the standard
solvers included in OpenFOAM without modification, i.e., it can be readily used
with RANS or LES, multiphase models (e.g., for simulating the free surface in
MHK installations), and even with heat transfer.

\section{Results}

\begin{figure}
    \centering

    \includegraphics[width=0.8\textwidth]{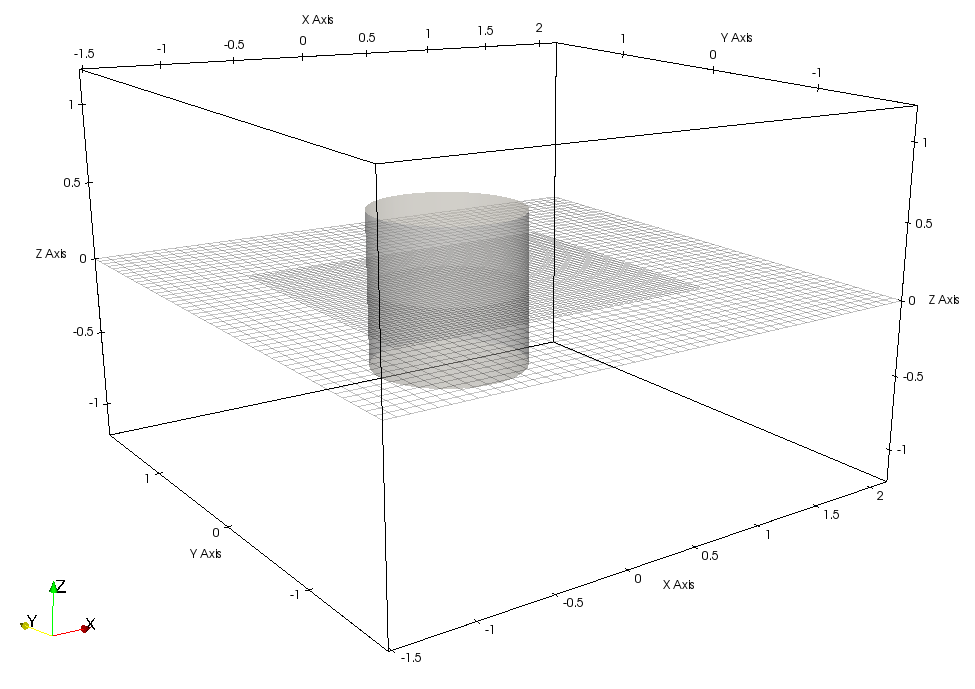}

    \caption{Computational domain used for the RANS ALM simulations. All
    dimensions are in meters, flow is oriented in the $+x$-direction, and the
    cylindrical region represents the swept area of the UNH-RVAT rotor. Note
    that increased mesh refinement was used for the LES simulations.}

    \label{fig:ALM-mesh}
\end{figure}

Both the high solidity UNH-RVAT and medium solidity RM2 turbines were modeled
using the ALM inside a Reynolds-averaged Navier--Stokes (RANS) simulation,
closed with the standard $k$--$\epsilon$ turbulence model. These rotors provide
diverse parameters, which helped evaluate the robustness of the ALM. The
simulations were performed inside a domain similar in size to that used in
\cite{Bachant2016-BR-CFD} to approximatel (including blockage) the tow tank
experiments (3.66 m wide, 2.44 m tall, 1.52 m upstream and 2.16 m downstream),
with similar velocity boundary conditions: 1 m/s inflow, no-slip (fixed to 1
m/s) walls and bottom, and a rigid slip condition for the top, to approximate
some effects of the tow tank's free surface. Simulations were run for a total of
6 seconds, with the latter half used to calculate performance and wake
statistics. Pressure--velocity coupling for the momentum equation was achieved
using the PISO (pressure implicit splitting of operators) method. A slice of the
mesh in the $x$--$y$ plane is shown in Figure~\ref{fig:ALM-mesh}.

Similar numerical settings were used for each turbine as well. The Sheng et al.
DS model was used with the default coefficients given in \cite{Sheng2008}, and
the Goude flow curvature correction was employed. A second order backward
difference was used for advancing the simulation in time, and second order
linear schemes were used for the majority of the terms' spatial discretizations.
The only major difference between the two simulation configurations was that the
end effects model was deactivated for the RM2, since it reduced $C_P$ far below
the experimental measurements. This modification is consistent with the RM2
blades' higher aspect ratio (15 versus the UNH-RVAT's 7.1) and tapered planform,
though will need to be investigated further. The number of elements per actuator
line was set to be approximately equal to the total span divided by the Gaussian
force projection width $\epsilon$. Case files for running all the simulations
presented here with OpenFOAM v5.0 are available from
\cite{Bachant2018-UNH-RVAT-turbinesFoam-v1.1.0,
Bachant2018-RM2-turbinesFoam-v1.1.0}.


\subsection{Verification}

\begin{figure}
    \centering

    \includegraphics[width=0.7\textwidth]{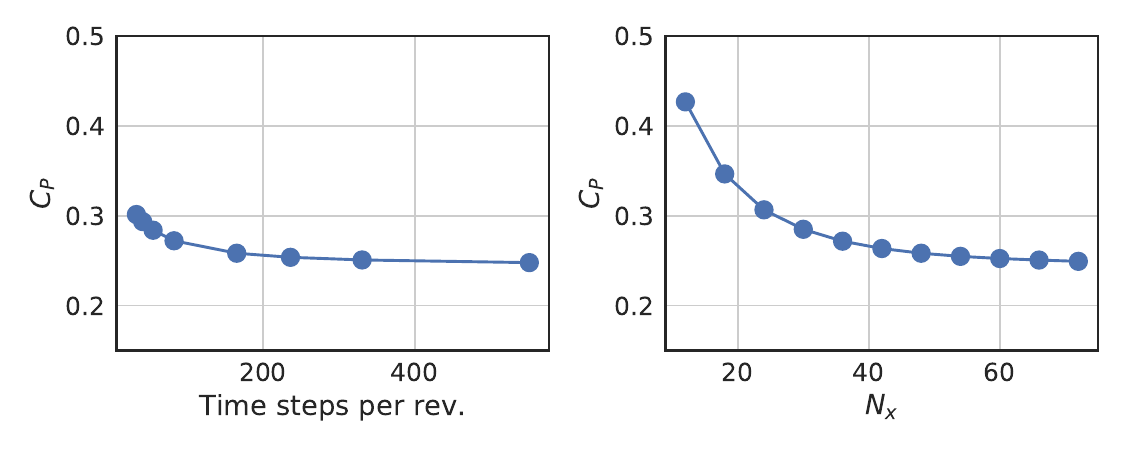}

    \caption{Temporal (left) and spatial (right) grid resolution sensitivity
        results for the UNH-RVAT ALM RANS model.}

    \label{fig:RVAT-ALM-verification}
\end{figure}

\begin{figure}
    \centering

    \includegraphics[width=0.7\textwidth]{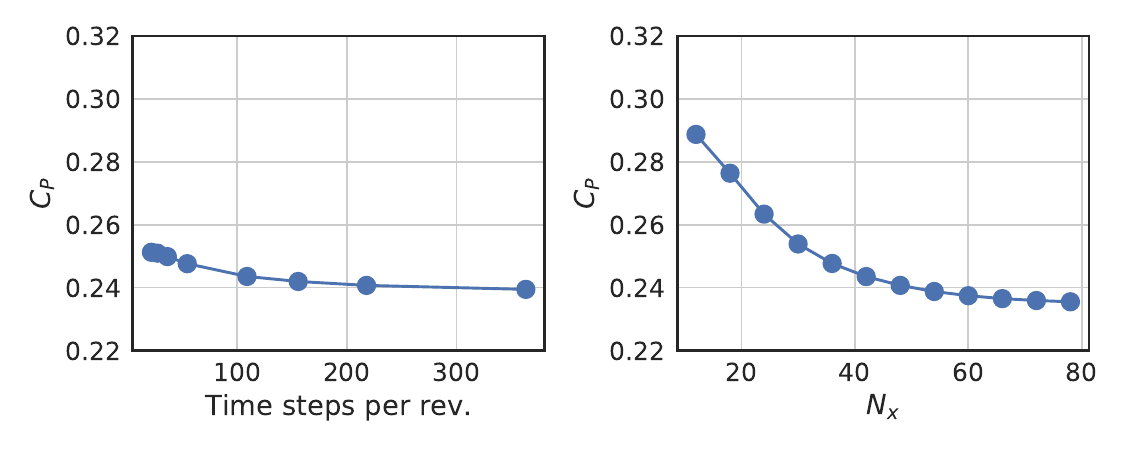}

    \caption{Temporal (left) and spatial (right) grid resolution sensitivity
        results for the RM2 ALM RANS model.}

    \label{fig:RM2-ALM-verification}
\end{figure}

Verification for sensitivity to spatial and temporal grid resolution was
performed for both the UNH-RVAT and RM2 RANS cases at their optimal tip speed
rations, the results from which are plotted in
Figure~\ref{fig:RVAT-ALM-verification} and
Figure~\ref{fig:RM2-ALM-verification}, respectively. Similar to the verification
strategy employed in \cite{Bachant2016-BR-CFD}, the mesh topology was kept
constant, and the resolution was scaled proportional to the number of cells in
the $x$-direction $N_x$ for the base hexahedral mesh, which was essentially
uniform in resolution in all directions. Both models displayed low
sensitivity to the number of time steps per revolution. Spatial grid dependence,
however, was more important.

Final spatial grid resolutions were chosen as $N_x=48$ for both the UNH-RVAT and
RM2 cases. Time steps were chosen as $\Delta t_0 = 0.01$ and
$\Delta t_0 = 0.005$
seconds for the UNH-RVAT and RM2 respectively, which correspond to approximately
200 steps per revolution. The chosen values should provide $C_P$ predictions
within one percentage point of the true solution, which is on the order of the
expanded uncertainty of the experimental measurements. Note that for computing
performance curves, the number of steps per revolution was kept constant, i.e.,
the time step was adjusted to $\Delta t = \Delta t_0 \lambda_0 / \lambda$,
where $\lambda_0$ was the tip speed ratio roughly corresponding to peak $C_P$
as observed in experiments, which was used for verifying the numerical
sensitivity to spatial and temporal resolution.

\subsection{UNH-RVAT RANS}

\begin{figure}
    \centering

    \includegraphics[width=0.7\textwidth]{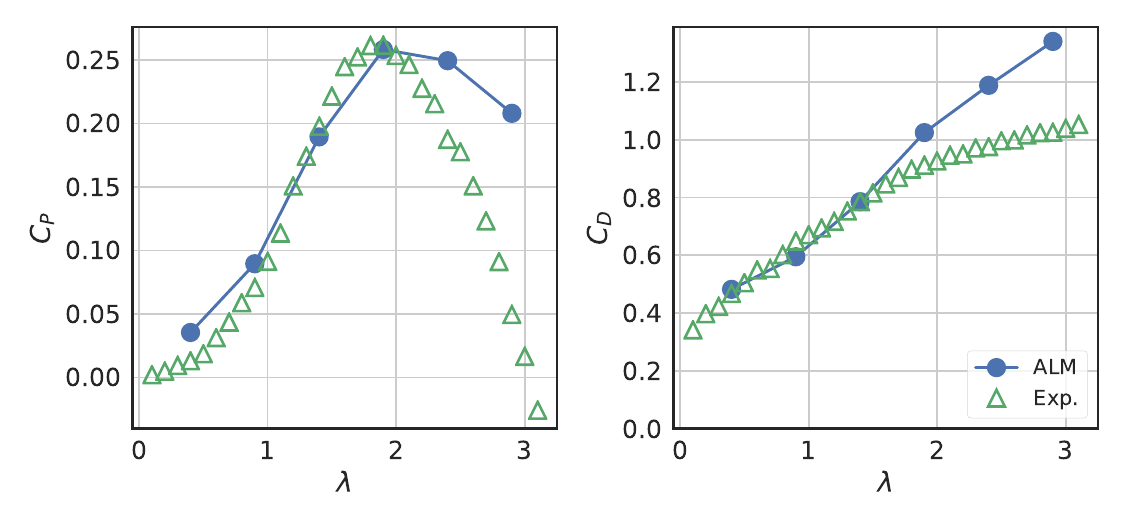}

    \caption{Power and drag coefficient curves computed for the UNH-RVAT using
        the actuator line model with RANS, compared with experimental results
        from~\cite{Bachant2016-RVAT-Re-dep}.}

    \label{fig:RVAT-ALM-perf-curves}
\end{figure}

Power and overall rotor drag (a.k.a. thrust) coefficient curves are plotted for
the UNH-RVAT in Figure~\ref{fig:RVAT-ALM-perf-curves}. The ALM was successful at
predicting the performance tip speed ratios up to $\lambda_0$, which suggests
that dynamic stall was being modeled accurately, but $C_P$ was overpredicted at
high $\lambda$. This may have been caused by the omission of additional
parasitic drag sources such as roughness from exposed bolt heads located far
enough from the axis to have a large effect at high rotation rates, or an
underestimation of the blade--strut connection corner drag coefficient. In
\cite{Rawlings2008,Bachant2016-RM2-paper} it was shown how these losses can be
significant even with carefully smoothed struts and strut-blade connections.
Overprediction of performance at high tip speed ratio could also be a
consequence of the Leishman--Beddoes dynamic stall model, which can also be seen
in the Darrieus VAWT momentum model results shown in Figure 6.70 of
\cite{Para2002}.


\begin{figure}
    \centering

    \includegraphics[width=0.75\textwidth]{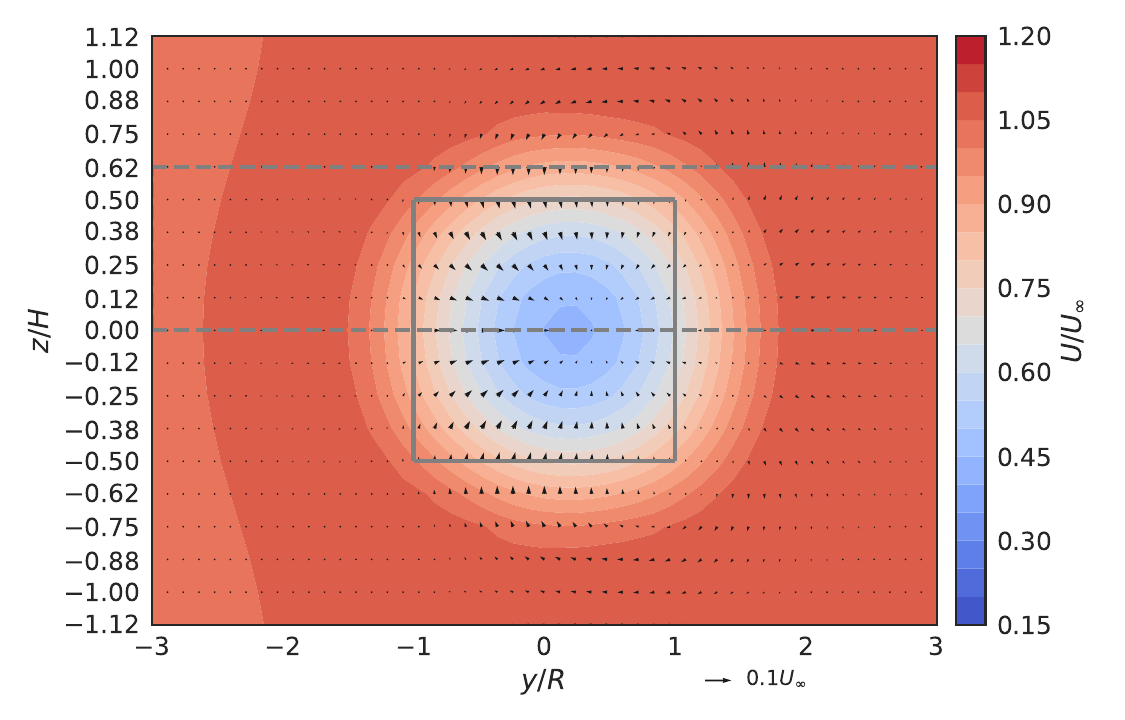}

    \caption{UNH-RVAT mean velocity field at $x/D=1$ computed with the ALM
        coupled with a $k$--$\epsilon$ RANS model.}

    \label{fig:RVAT-ALM-meancontquiv}
\end{figure}

Figure~\ref{fig:RVAT-ALM-meancontquiv} shows mean velocity field for the
UNH-RVAT computed by the ALM RANS model. The asymmetry observed in the
experiments \cite{Bachant2015-JoT} was captured well, along with some of the
vertical flow due to blade tip vortex shedding, though the flow structure is
missing the detail present in the experiments and blade-resolved RANS
simulations. Overall, the wake appears to be over-diffused, which could be a
consequence of the relatively coarse mesh. Note that with the DS and flow
curvature corrections turned off, the direction of the mean swirling motion
reverses, which highlights the importance of resolving the correct azimuthal
location of blade loading.

\begin{figure}
    \centering

    \includegraphics[width=0.7\textwidth]{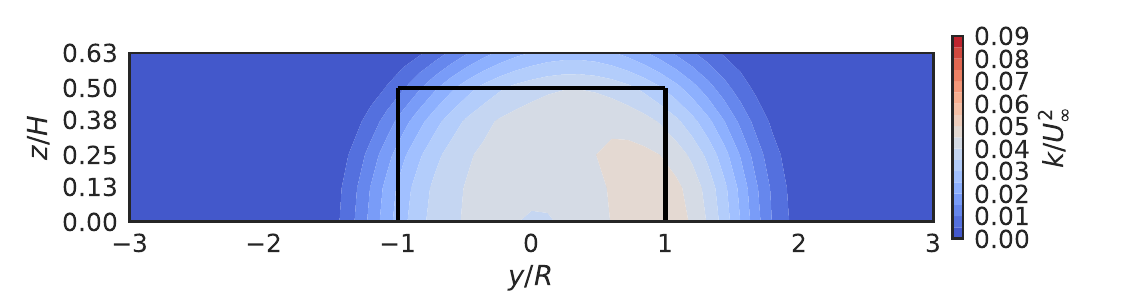}

    \caption{UNH-RVAT turbulence kinetic energy contours at $x/D=1$ predicted by
        the ALM inside a $k$--$\epsilon$ RANS model.}

    \label{fig:RVAT-ALM-kcont}
\end{figure}

Turbulence kinetic energy contours (including resolved and modeled energy) are
shown in Figure~\ref{fig:RVAT-ALM-kcont}. The ALM was able to resolve the
concentrated area of $k$ on the $+y$ side of the turbine, present in the
experiments~\cite{Bachant2015-JoT}, but the turbulence generated by the dynamic
stall vortex shedding process is absent. This makes sense since in the ALM, the
DS model only modulates the body force term in the momentum equation, which does
not provide a mechanism for mimicking shed vortices or turbulence.

\begin{figure}
    \centering

    \includegraphics[width=0.7\textwidth]{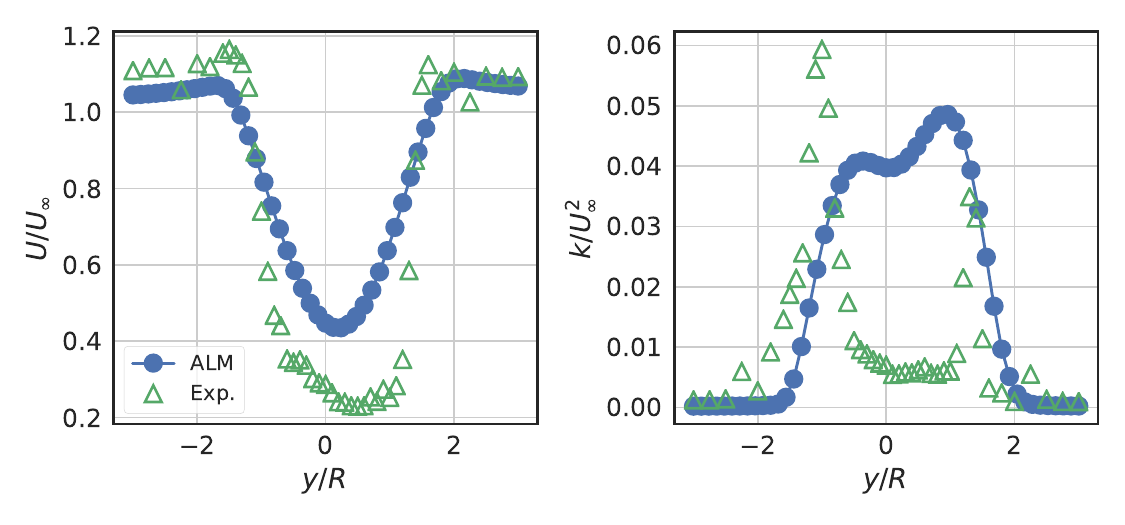}

    \caption{Mean streamwise velocity (left) and turbulence kinetic energy
        (right) profiles at $z/H=0$ for the UNH-RVAT ALM, compared with the
        experimental data from~\cite{Bachant2016-RVAT-Re-dep}.}

    \label{fig:RVAT-ALM-profiles}
\end{figure}

Profiles of mean streamwise velocity and turbulence kinetic energy are shown in
Figure~\ref{fig:RVAT-ALM-profiles}. Here the over-diffused or over-recovered
characteristic of the mean velocity deficit seen in
Figure~\ref{fig:RVAT-ALM-meancontquiv} is more apparent. This effect is also
seen in the profile of $k$, where energy is smeared over the center region of
the rotor.

\begin{figure}
    \centering

    \includegraphics[width=0.7\textwidth]{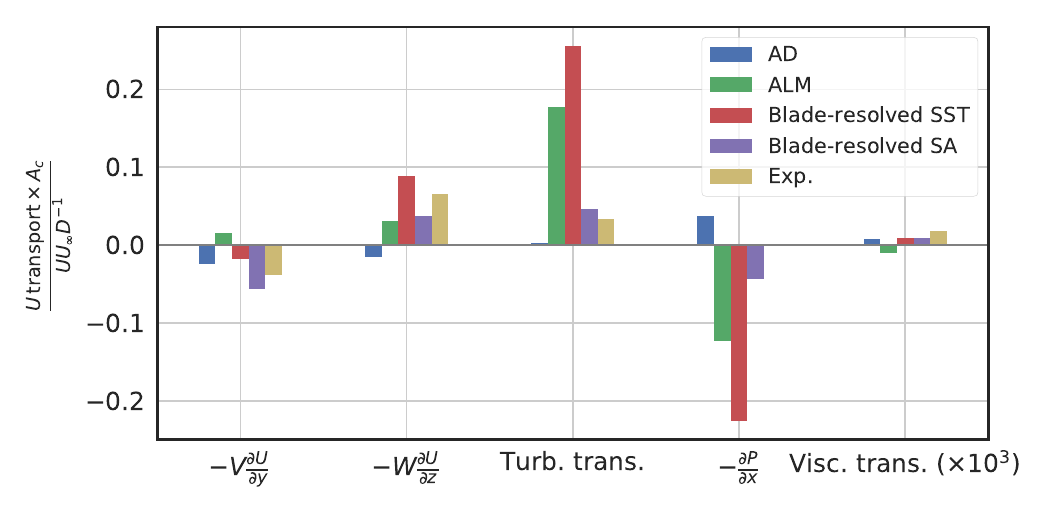}

    \caption{
        Weighted average momentum recovery terms for the actuator disk (AD)
        simulation from \cite{Bachant2015-JoT}, UNH-RVAT actuator line model
        with a $k$--$\epsilon$ RANS closure, the two 3-D blade resolved RANS
        models described in \cite{Bachant2016-BR-CFD} ($k$--$\omega$ SST and
        Spalart--Allmaras, SA), and the experiments reported in
        \cite{Bachant2015-JoT}.
    }

    \label{fig:RVAT-ALM-recovery}
\end{figure}

Weighted averages for the terms in the streamwise momentum equation were
computed identically as they were in \cite{Bachant2015-JoT,Bachant2016-BR-CFD},
and are plotted in Figure~\ref{fig:RVAT-ALM-recovery} along with the actuator
disk (AD) results from \cite{Bachant2015-JoT}, 3-D blade-resolved RANS results
from \cite{Bachant2016-BR-CFD} and experiments. The most glaring discrepancy is
the ALM's prediction of positive cross-stream advection, which is caused by the
lack of detail in the tip vortex shedding. The total for vertical advection,
however, is close to that predicted by the 3-D blade-resolved Spalart--Allmaras
model. Levels of turbulent transport due to eddy viscosity and deceleration due
to the adverse pressure gradient are between those predicted by the 3-D
blade-resolved $k$--$\omega$ SST and SA models. Overall, however, one might
expect the total wake recovery rate to be comparable between all models except
the actuator disk, which induces negative vertical advection, very little
turbulent transport, and has a positive pressure gradient contribution. These
results suggests the ALM would be an effective tool---much better than an
actuator disk---for assessing downstream spacing of subsequent CFTs, though
blade--vortex interaction for very tightly spaced rotors may not be captured, at
least on relatively coarse meshes as used here.

\subsection{RM2 RANS}

\begin{figure}
    \centering

    \includegraphics[width=0.7\textwidth]{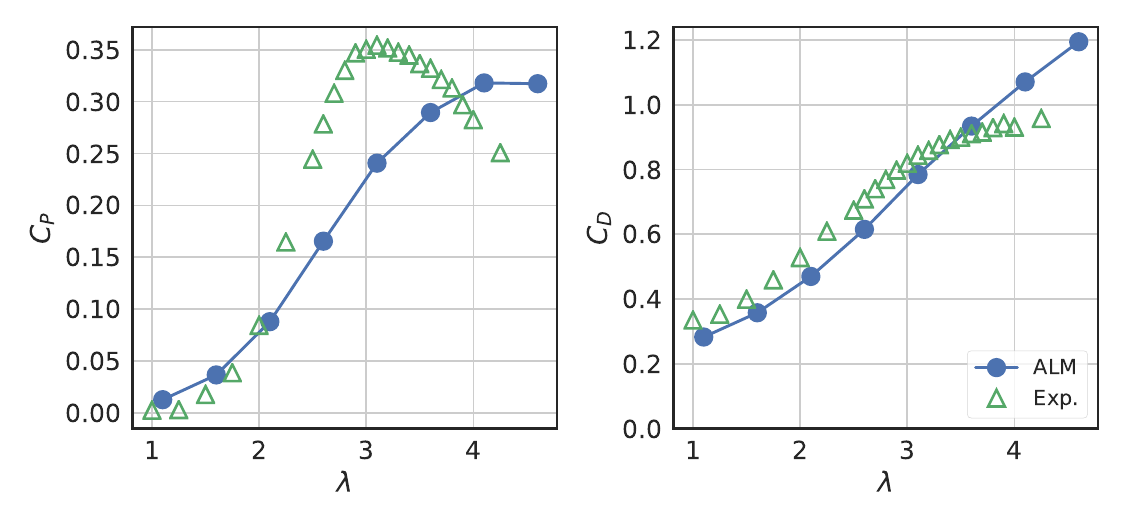}

    \caption{Power and drag coefficient curves computed for the RM2 using the
        ALM, compared with experimental data from~\cite{Bachant2016-RM2-data}.}

    \label{fig:RM2-ALM-perf-curves}
\end{figure}

Figure~\ref{fig:RM2-ALM-perf-curves} shows the performance curves computed for
the RM2 by the ALM, and those from the tow tank
experiments~\cite{Bachant2016-RM2-paper}. As with the high solidity RVAT, $C_P$
is overpredicted at high $\lambda$. However, $\lambda_0$, the tip speed ratio of
peak power coefficient, is also shifted to the right. This is indicative of
inaccurate dynamic stall modeling, which could possibly be attributed to one of
the models' tuning constants, e.g., the time constant $T_\alpha$ for the lagged
angle of attack. Limited ad hoc testing revealed that the mean $C_P$ at
$\lambda_0$ more closely matched experimental measurements with $T_\alpha$
roughly double the default value given in~\cite{Sheng2008}. This could be
investigated further by looking at the phase of, e.g., maximum $C_P$ peaks
compared to the experimental values (or possibly those from blade-resolved CFD),
but for the present study only mean values were considered.


\begin{figure}
    \centering

    \includegraphics[width=0.8\textwidth]{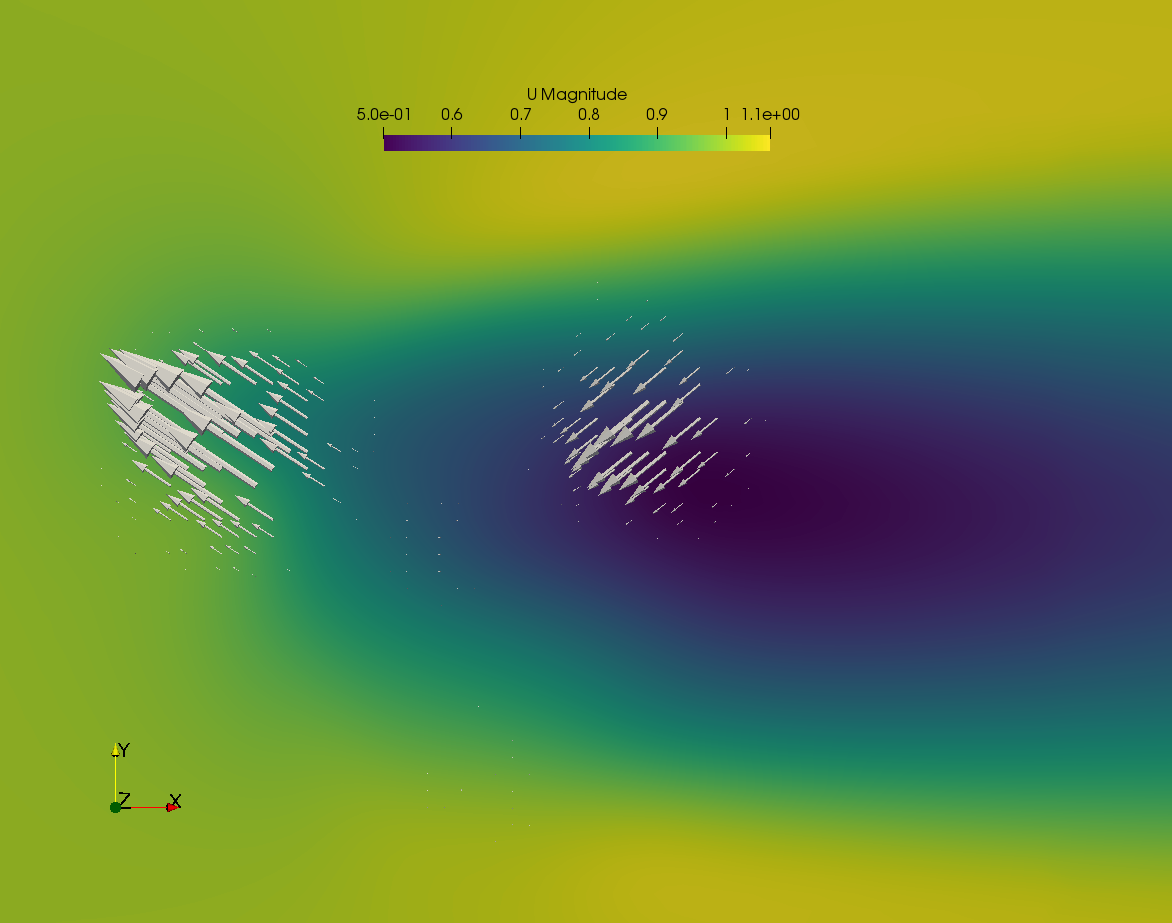}

    \caption{Velocity magnitude and actuator line force field vectors for the
    RM2 RANS simulation at 6 seconds simulated time.}

    \label{fig:RM2-internal}
\end{figure}

The internal rotor flow about the horizontal center plane, along with
the directions and relative magnitudes of the actuator line force vectors
are shown in Figure~\ref{fig:RM2-internal}.
It can be seen how flow features like dynamic stall vortices are not
captured with the ALM, despite forces themselves being modulated by the
dynamic stall model. This combined with the diffusive nature of the
RANS model helps explain discrepancies in predicted power coefficient.
With the computational affordability of the ALM coupled with RANS comes a
compromise in flow detail, which should be acceptable for array-level
simulation, but for individual device design a blade-resolved
simulation would be more advisable.

The mean near-wake structure for the RM2 looked qualitatively similar to the
RVAT ALM RANS case, but for the RM2, the effects of blade tip vortex shedding
were weaker, which is consistent with experiments~\cite{Bachant2016-RM2-paper}.
Nonetheless, the mean vertical flow towards the $x$--$y$ center plane was
captured, which is an important qualitative feature of both CFT near-wakes.
Like for the UNH-RVAT, predicted turbulence kinetic energy values were
concentrated on the $+y$ side of the rotor. However, overall levels of
turbulence are lower than for the UNH-RVAT, which is also consistent with the
experimental results.

As for the UNH-RVAT RANS case, a mean streamwise momentum transport analysis was
undertaken for the RM2 near-wake by computing weighted sums of each term across
the entire domain in the $y$--$z$ directions. Similar results as for the
UNH-RVAT were obtained, i.e., cross-stream advection was predicted to be
positive where it should have been negative, vertical advection was predicted
reasonably well, and turbulent transport due to the eddy viscosity was also
relatively large. The ratio of wake transport compared with the UNH-RVAT RANS
case (approximately 60\% to 70\% lower for the RM2) matches well with that
computed from the experiments, which shows the ALM may successfully predict
larger optimal array spacing for the RM2 versus UNH-RVAT.

\subsection{UNH-RVAT LES}

The state-of-the-art in high fidelity turbine array modeling uses the actuator
line method coupled with large eddy simulation (LES), which allows more of the
turbulent energy spectrum to be directly resolved, only requiring the dynamics
of the smallest scales---where dissipation occurs---to be computed by the
so-called subgrid-scale (SGS) model. Since the ALM LES approach has only been
reported in the literature for a very low Reynolds number 2-D CFT
\cite{Shamsoddin2014}, and CFTs may provide unique opportunities to array
optimization, which could be explored with LES, it was of interest to determine
how well the ALM coupled with LES might predict wake dynamics of a higher $Re$
3-D CFT rotor.

Thus, the UNH-RVAT baseline case was simulated using the Smagorinsky LES
turbulence model \cite{Smagorinsky1963}, which was the first of its kind, and
serves as a good baseline for LES modeling since its behavior is well-reported
in the literature, especially for ALM simulations. OpenFOAM's default
Smagorinsky model coefficients were used (giving an approximate equivalent
Smagorinsky coefficient $C_S = 0.17$), and the LES filter width was set as the
cube root of the local cell volume. The tip speed ratio was set to oscillate
sinusoidally about $\lambda_0$ with a 0.19 magnitude and the angle of the first
peak at 1.4 radians---similar to the rotation presribed in the blade-resolved
RANS simulations discussed in \cite{Bachant2016-BR-CFD}.

Since the computational cost of LES is significantly higher than RANS,
verification with respect to grid dependence was not performed. Instead, mesh
resolution was chosen relative to similar studies of turbine wake ALM LES. Of
the studies surveyed
\cite{Shamsoddin2014,Archer2013,Martinez-Tossas2015a,Troldborg2007}, the mesh
resolution ranged from 18--64 points per turbine diameter. The mesh here was set
accordingly by using a 16 point per meter base mesh, and refining twice in a
region containing the turbine to produce a 64 point per turbine diameter/height
resolution. The solver was run with a 0.002 second time step, which is
significantly within the limit described by \cite{Martinez-Tossas2015}, where an
actuator line element may not pass through more than one cell per time step.
With these resolutions computation times were $O(10)$ CPU hours per second of
simulated time (on 4 processors),
which was approximately two orders of magnitude lower than for
blade-resolved RANS.

\begin{figure}
    \centering

    \includegraphics[width=0.8\textwidth]{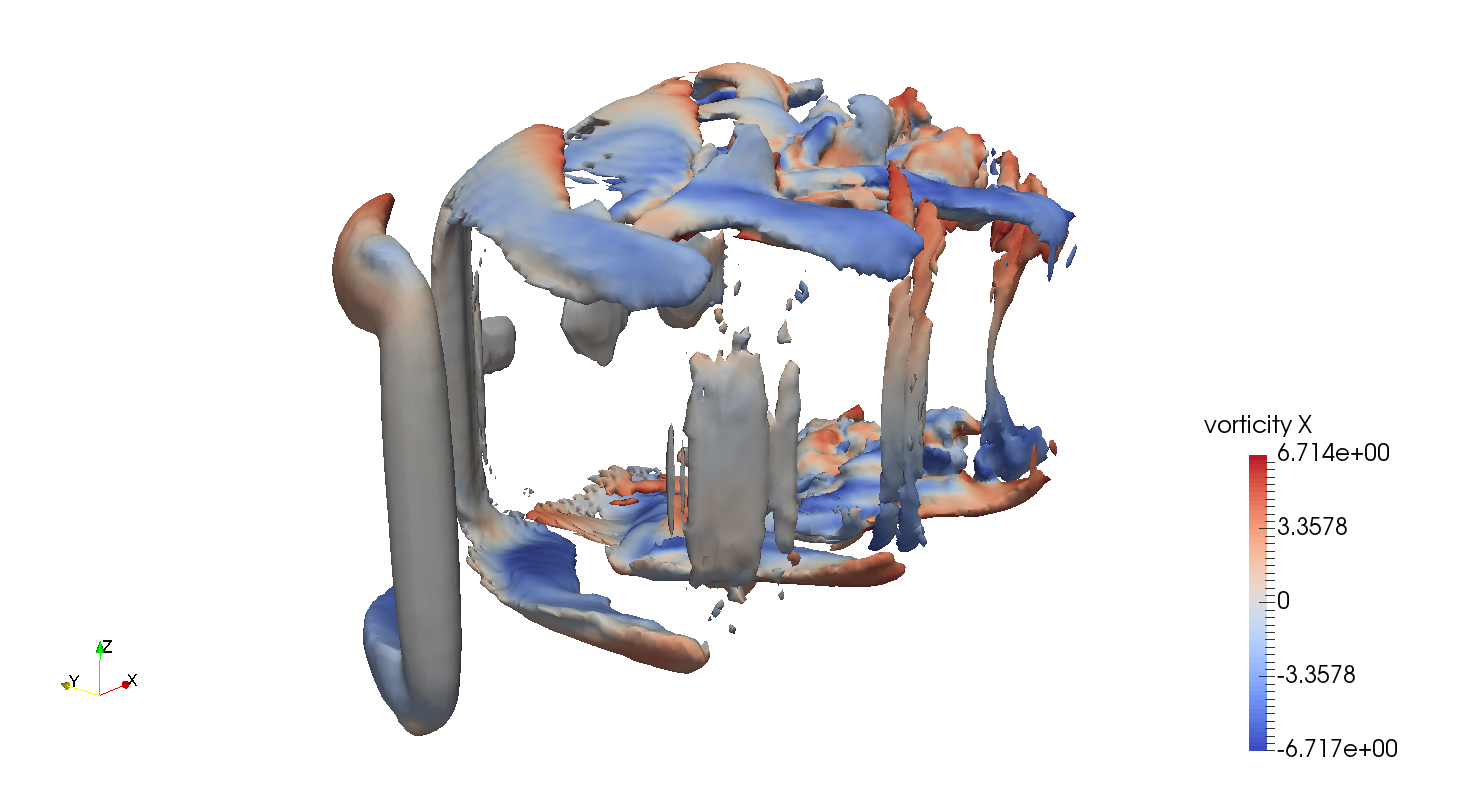}

    \caption{Snapshot of vorticity isosurfaces (colored by their streamwise
        component) at $t=6$ s for the UNH-RVAT LES case.}

    \label{RVAT-ALM-LES-vorticity}
\end{figure}

Mean power coefficient predictions for the UNH-RVAT at its optimal mean tip
speed ratio dropped to 0.20 using the ALM within the large eddy simulation.
However, the amount of information regarding the wake dynamics was greatly
increased, even beyond that of the 3-D blade-resolved RANS.
Figure~\ref{RVAT-ALM-LES-vorticity} shows an instantaneous snapshot of
isosurfaces of vorticity produced by the actuator lines.

\begin{figure}
    \centering

    \includegraphics[width=0.75\textwidth]{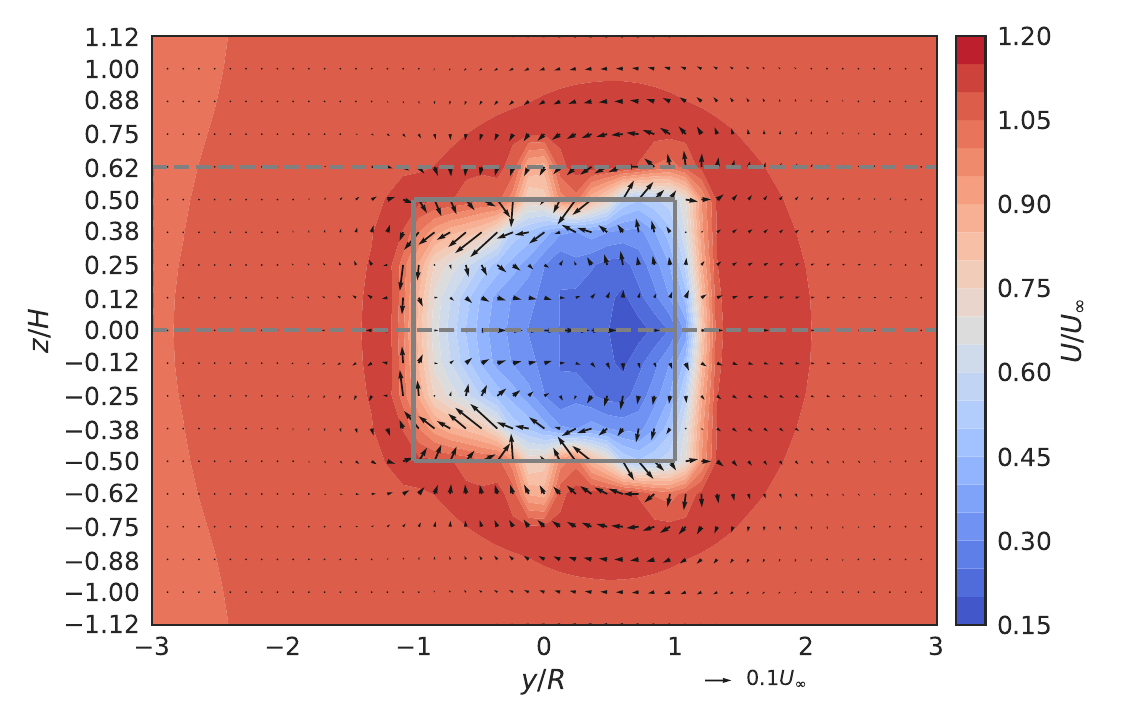}

    \caption{Mean velocity field in the UNH-RVAT near-wake at $x/D=1$ computed
        with the Smagorinsky LES model.}

    \label{fig:RVAT-ALM-LES-meancontquiv}
\end{figure}

The near-wake's mean velocity field at $x/D=1$ is shown in
Figure~\ref{fig:RVAT-ALM-LES-meancontquiv}. Compared with the RANS ALM results,
the LES looks much more like the blade-resolved and experimental results
\cite{Bachant2015-JoT}, showing the clockwise and counterclockwise mean swirling
motion on the $-y$ and $+y$ sides of the rotor, respectively.

\begin{figure}
    \centering

    \includegraphics[width=0.7\textwidth]{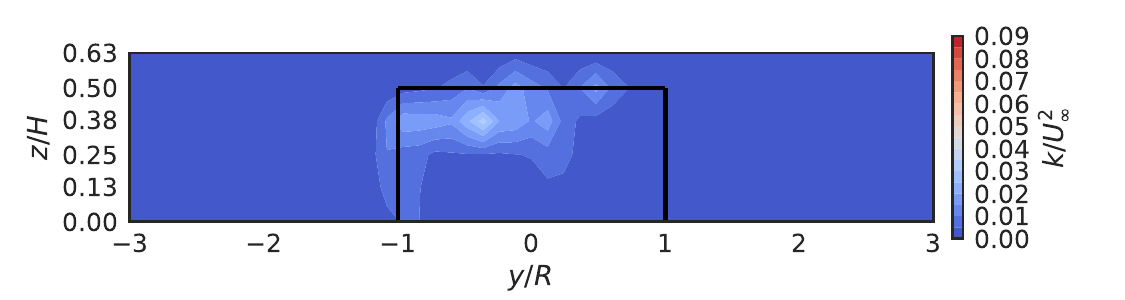}

    \caption{Turbulence kinetic energy in the UNH-RVAT near-wake at $x/D=1$
        computed with the Smagorinsky LES model.}

    \label{fig:RVAT-ALM-LES-kcont}
\end{figure}

Contours of turbulence kinetic energy sampled at $x/D=1$ from the large eddy
simulation are plotted in Figure~\ref{fig:RVAT-ALM-LES-kcont}. Compared with
RANS, LES is more able to predict the turbulence generated by the blade tip
vortex shedding and dynamic stall effects, though the total turbulence kinetic
energy was lower, especially on the $+y$ side of the rotor.
This is likely a consequence of the SGS modeling, where the vortical structures
generated by the blades remain stable further downstream. Similar effects were
seen in \cite{Martinez-Tossas2015a, Shamsoddin2014}, where higher levels of the
Smagorinsky coefficient delayed vortex breakdown and subsequent higher levels of
turbulence. Figure~\ref{RVAT-ALM-LES-vorticity} shows evidence of these effects,
where the blade bound and tip vortices are still relatively coherent at $x/D=1$.

\begin{figure}
    \centering

    \includegraphics[width=0.7\textwidth]{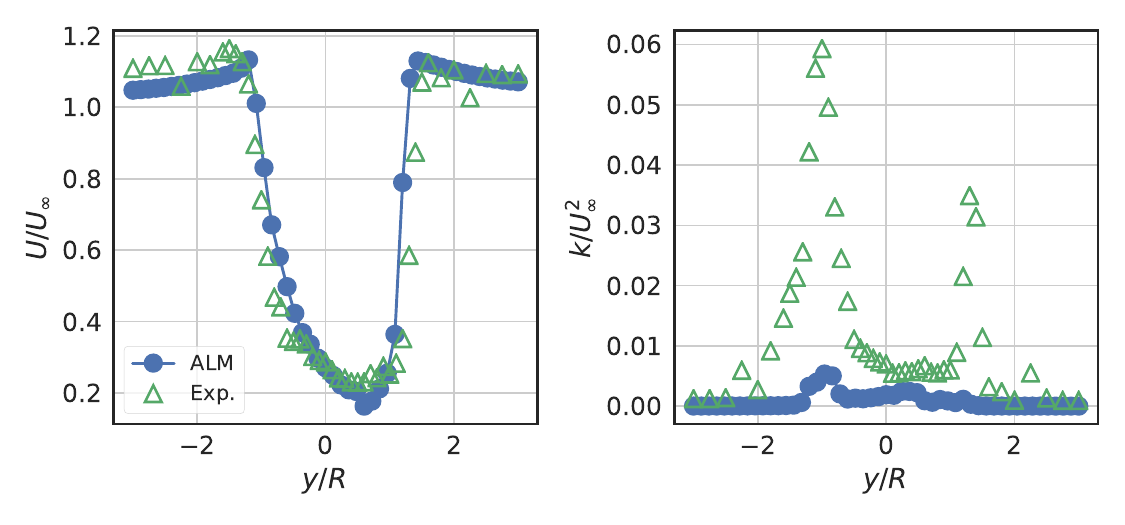}

    \caption{Mean velocity profiles in the UNH-RVAT near-wake at $x/D=1$ and
        $z/H=0$ computed with the Smagorinsky LES model, compared with experimental
        data from~\cite{Bachant2016-RVAT-Re-dep}.}

    \label{fig:RVAT-ALM-LES-profiles}
\end{figure}

Mean velocity profiles at the turbine center plane, plotted in
Figure~\ref{fig:RVAT-ALM-LES-profiles}, were predicted more accurately using LES
versus RANS, and rival those of the 3-D blade-resolved models. However,
turbulence kinetic energy profiles did not match as closely with experiments.
Though the qualitative shape was resolved better than that by the RANS ALM
simulation, notably the asymmetric peaks around $y/R = \pm 1$, the turbulence
generated in the large eddy simulation was approximately an order of magnitude
too low.
Note that $k$ (both resolved and SGS) does increase significantly downstream
after the transition in grid resolution, which implies the grid can help
compensate for the artificial stability of the vortex structures caused by
the Smagorinsky model.
It is hypothesized that the turbulence kinetic energy would better match
experimental data taken further downstream, meaning the model should perform
well for array simulations with spacing as close as $2D$.

The planar weighted sums of streamwise momentum recovery terms were computed in
the same way as for the RANS cases with the exception of the turbulent transport
term, which for the LES was computed from the $x$-components of the divergence
of the resolved and SGS Reynolds stress tensors:
\begin{equation}
    \text{Turb. trans.} = - \left( \frac{\partial}{\partial x_j}
    \overline{u^\prime_x u^\prime_j}
    + \frac{\partial}{\partial x_j} R_{xj}
    \right),
\end{equation}
where $u$ indicates the resolved or filtered velocity, and $R$ is the
subgrid-scale Reynolds stress.

\begin{figure}
    \centering

    \includegraphics[width=0.7\textwidth]{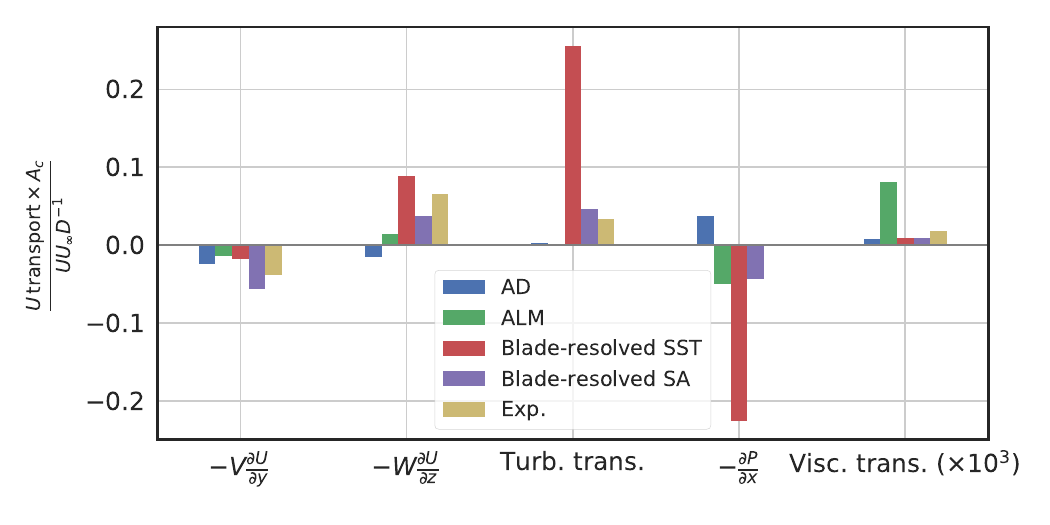}

    \caption{Weighted average momentum recovery terms at $x/D=1$ for the RVAT
        ALM LES using the Smagorinsky SGS model.}

    \label{fig:RVAT-ALM-LES-recovery}
\end{figure}

Transport term weighted sums computed from the LES results are shown in
Figure~\ref{fig:RVAT-ALM-LES-recovery}. Unlike the RANS ALM cases, the
cross-stream advection contributions are negative, as they are in the experiment
and blade-resolved CFD models. The vertical advection term is positive as
expected, though smaller than in other cases. Interestingly, the turbulent
transport is negative in the LES, meaning the combined effects of the resolved
and SGS stressed are transferring momentum out of the wake. The low turbulent
transport appears to be partially balanced by higher levels of viscous
diffusion---about an order of magnitude larger than the 3-D blade-resolved RANS
models and experiments. These discrepancies highlight the difficulty of
predicting the near-wake dynamics, the importance of the SGS model in LES, and
the need for data further downstream to test and refine predictions for wake
evolution. For example, setting the Smagorinsky coefficient higher may induce
vortex breakdown earlier, which would raise the turbulence levels significantly.

\section{Conclusions}

An actuator line model for cross-flow turbines, including a Leishman--Beddoes
type dynamic stall model, flow curvature, added mass, and lifting-line based end
effects corrections, was developed and validated against experimental datasets
acquired for high and medium solidity rotors at scales where the performance and
near-wake dynamics were essentially Reynolds number independent. When coupled to
a $k$--$\epsilon$ RANS solver ALM simulations took $O(0.1)$ CPU hours per second
of simulated time, while when coupled with a Smagorinsky LES model the computing
time was $O(10)$ hours per second, which represent a four and two order of
magnitude decrease in computational expense versus 3-D blade-resolved
RANS~\cite{Bachant2016-BR-CFD}, respectively.

The RANS ALM predicted the UNH-RVAT performance well at tip speed ratios up to
and including that of max power coefficient. The RM2 power coefficient on the
other hand was underpredicted at lower $\lambda$. Both models overestimated
$C_P$ at the highest tip speed ratios, which has been observed in other
simulations using Leishman--Beddoes type dynamic stall models. Possible
explanations include underestimation of added mass effects or blade--strut
connection corner drag, incorrect time constants in the LB DS model, and/or
inaccuracy due to the virtual camber effect. In the present flow curvature
model, the angle of attack is corrected, but the foil coefficient data is not
transformed, meaning the LB DS separation point curve fit parameters are equal
for both positive and negative angles of attack. A foil data transformation
algorithm based on virtual camber should be investigated for future improvement
of the ALM.

The RANS ALM cases were able to match some important qualitative near-wake flow
features, e.g., the mean vertical advection velocity towards the mid-rotor
plane. However, the mean flow structure and turbulence generation due to blade
tip and dynamic stall vortex shedding shows some discrepancy with experimental
and blade-resolved CFD. Extensions to the ALM to deal with these shortcomings
should be developed, e.g., a turbulence injection model as employed by James et
al. \cite{James2010} or a model that will ``turn'' the ALM body force vectors to
approach the effects of leading and trailing edge vortex shedding during dynamic
stall.

The UNH-RVAT was simulated with the ALM embedded within a typical Smagorinsky
LES, which thanks to its lower diffusion and/or dissipation was able to more
accurately capture the large scale vortical flow structures shed by the rotor
blades. Turbulence generated by the blade tip vortex shedding and dynamic stall
region of the blade path was better resolved, but overall lower levels of
turbulence were predicted, which is likely a consequence of the subgrid-scale
model's influence on the stability of shed vortices. This effect was also
apparent in the negative predictions of turbulent transport on the streamwise
momentum recovery. Therefore, subgrid-scale modeling should be investigated
further before applying the ALM LES to array analyses.

The ALM provides a more physical flow description compared to momentum and
potential flow vortex models, at a reasonable cost. The ALM also drastically
reduces computational effort compared to blade-resolved CFD, while maintaining
the unsteadiness of the wake not resolved by a conventional actuator disk. When
combined with RANS the ALM will allow VAT array simulations on individual PCs,
and with high performance computing and LES the ALM is one of the highest
fidelity array modeling tools available. Ultimately, the ALM will help reduce
dependence on expensive physical modeling to optimize VAT array layouts.

\acknowledgments{
    The authors would like to acknowledge funding through a National Science
    Foundation CAREER award (principal investigator Martin Wosnik, NSF 1150797,
    Energy for Sustainability, program manager Gregory L. Rorrer). This work has
    been carried out as part of author P. Bachant's doctoral research at the
    University of New Hampshire.
}

\authorcontributions{
    P.B. wrote the computer code, ran the simulations, post-processed the
    results, and wrote the paper.
    A.G. contributed to the computer code and the manuscript.
    M.W. contributed to the writing and editing of the manuscript.
}

\conflictsofinterest{The authors declare no conflicts of interest.}



\externalbibliography{yes}
\bibliography{library}

\end{document}